\documentclass[%
 reprint,
superscriptaddress,
 amsmath,amssymb,
 aps,
]{revtex4-1}

\usepackage{graphicx}% Include figure files
\usepackage{dcolumn}% Align table columns on decimal point
\usepackage{bm}% bold math
\usepackage{hyperref}% add hypertext capabilities
\begin{document}
%\preprint{APS/123-QED}

\title{Combining Laue diffraction with Bragg coherent diffraction imaging at 34-ID-C}% Force line breaks with \\

\author{Anastasios Pateras}
\email{aptrs@lanl.gov}
\affiliation{Materials Science and Technology Division, Los Alamos National Laboratory, Los Alamos, NM 87545, USA}%Lines break automatically or can be forced with \\
\author{Ross Harder}%
\email{rharder@anl.gov}
\author{Wonsuk Cha}
\author{Jon Tischler}
\author{Ruqing Xu}
\author{Wenjun Liu}
\author{Mark J. Erdmann}
\author{Robert Kalt}
\affiliation{Advanced Photon Source, Argonne National Laboratory, Lemont, IL 60439, USA}
\author{Jonathan G. Gigax}
\author{J. Kevin Baldwin}
\affiliation{Center for Integrated Nanotechnologies, Los Alamos National Laboratory, Los Alamos, NM 87545, USA}
\author{Richard L. Sandberg}
\affiliation{Department of Physics and Astronomy, Brigham Young University, Provo, UT 84602, USA}
\author{Saryu Fensin}
\author{Reeju Pokharel}
\affiliation{Materials Science and Technology Division, Los Alamos National Laboratory, Los Alamos, NM 87545, USA}

\date{\today}% It is always \today, today,

\begin{abstract}
Measurement modalities in Bragg coherent diffraction imaging (BCDI) rely on finding signal from a single nanoscale crystal object, which satisfies the Bragg condition among a large number of arbitrarily oriented nanocrystals. However, even when the signal from a single Bragg reflection with (hkl) Miller indices is found, the crystallographic axes on the retrieved three-dimensional (3D) image of the crystal remain unknown, and thus, localizing in reciprocal space other Bragg reflections becomes in reality impossible or requires good knowledge of the orientation of the crystal. We report the commissioning of a movable double-bounce Si (111) monochromator at the 34-ID-C end station of the Advanced Photon Source, which aims at delivering multi-reflection BCDI as a standard tool in a single beamline instrument. The new instrument enables this through rapid switching from monochromatic to broadband (pink) beam permitting the use of Laue diffraction to determine crystal orientation. With a proper orientation matrix determined for the lattice, one can measure coherent diffraction near multiple Bragg peaks, thus providing sufficient information to image the full strain tensor in 3D. We discuss the design, concept of operation, the developed procedures for indexing Laue patterns, and automated measuring of Bragg coherent diffraction data from multiple reflections of the same nanocrystal. 
\end{abstract}

\maketitle

%\tableofcontents

\section{\label{sec:intro}Introduction}

Macroscopic properties of crystalline materials depend on their atomic structure, dimensionality, and other nanoscale phenomena. Dislocation locking, for example, is considered an effective physical mechanism for boost the hardness of polycrystalline metals \cite{RN3043}. Crystal slip has been shown to enhance surface diffusional creep and lead to unusual superplastic behavior of silver nanocrystals \cite{zhong_slip-activated_2017}. Elastic constants are typically defined with respect to bulk while their values deviate and need to be measured explicitly in the case of two-dimensional materials or nanometer-sized objects \cite{RN3021, RN3022}. Particularly in really small dimensions, the knowledge of the crystallographic orientation and full strain tensor are important pieces of information for predicting the mechanical properties of sub-micron particles and crystal grains of polycrystalline materials \cite{RN3025}. Bragg coherent x-ray diffraction imaging (BCDI) allows the visualization of the local atomic lattice displacement of single nanoparticles or grains in 3D. BCDI is compatible with \textit{in operando} measurements under different external stimuli, such as compression or tension, femtosecond laser light pulses, electric and magnetic fields allowing the visualization of evolving strain inside nanoparticles and ultimately the investigation of materials properties at the nanoscale \cite{RN624, RN2973, RN996, RN108, RN3004, RN410, RN32}. 

A frequent challenge faced in BCDI experiments, is that it relies on satisfying the Bragg condition of a single crystalline grain within a potentially large population of micron scale crystals or grains with unknown crystallographic orientations. As a result, it is often extremely difficult to measure more than one Bragg peak, thus, it is impossible to obtain all the components of the strain tensor \cite{RN1676}. To date, the strategies for collecting multi-reflection BCDI data have been to obtain important information about the crystallographic orientation of an isolated specimen before a BCDI experiment from other techniques, either scanning extensive volumes of reciprocal space until a reflection is found, or by preforming Laue diffraction beforehand at a different instrument \cite{RN3019, RN3015, RN1676}. 

A broadband (pink) x-ray beam permits Laue diffraction patterns to be measured from a lattice. The pattern of reflections arising on a detector is then a direct finger print of the orientation of the crystal lattice, assuming one knows the structure of the unit cell of the crystal \cite{RN3026, RN3027}. In this way, one can determine the crystallographic orientations of arbitrarily oriented crystals \cite{RN687, RN1017}. 

Until recently, it was not possible to index crystal grains in this way at 34-ID-C. Previous BCDI experiments relied on the sample containing many thousands of randomly oriented crystals, from which a signal would be detected largely by luck with the detector at a given position at a specified scattering angle for the lattice of interest. Here we present a new movable monochromator, recently commissioned at the APS 34-ID-C beamline, which allows a user to easily switch between a monochromatic and a pink x-ray beam. We also present the developed procedures and capabilities that allow multi-reflection BCDI at a single beamline instrument. The concept of operation relies on obtaining the crystallographic orientations of arbitrarily oriented sub-micron crystals utilizing a broadband x ray beam for Laue diffraction and a monochromatic beam for BCDI from different Bragg reflections \cite{RN3023}. The indexation of Laue patterns provides a detailed map of the crystal reflections in reciprocal space, assisting the localization of a desired Bragg peak. Then, by collecting at least three reflections, the three-dimensional (3D) image of the strain tensor of a nanocrystal can be obtained. This unique capability will be crucial for investigating properties of crystalline materials where the knowledge of the crystallographic orientation with respect to the axis of external stimuli is imperative \cite{RN1676, RN3004}.

\section{\label{sec:level1}Experimental section}
\subsection{\label{sec:level2}Design of the monochromator}

The 34-ID beamline at the Advanced Photon Source supports two experimental stations, a dedicated Laue diffraction microscopy instrument (34-ID-E) and the dedicated Coherent Diffraction Imaging instrument (34-ID-C). The 34-ID sector is fed by canted undulators with a 1 mrad angular separation between the beams. 
\begin{figure}[ht]
	\includegraphics[width=0.4\textwidth]{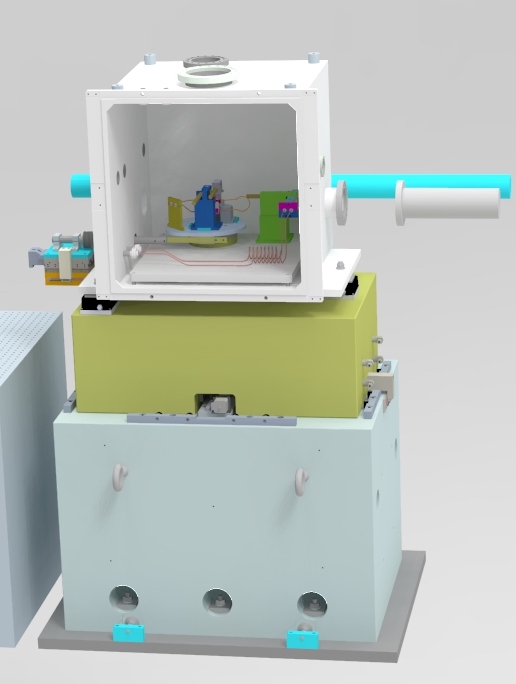}
	\caption{Front view of the monochromator as it is placed at 34-ID-C end station. The broadband x-ray beam is entering the vacuum chamber from the right, with the scattering plane being horizontal (Si crystals are vertical).}
\end{figure}
34-ID-C being the upstream instrument has a shielded transport running through the station with a 300 mm separation between the beams. This separation is gained by both the cant of the sources and a flat, cryogenically cooled, platinum coated mirror set at 5 mrad in the first optical enclosure of the beamline. The mirror filters the highest energies of the beamline’s 3.0 cm-period undulator providing x-rays up to roughly 17 keV at the end-station, with significantly lower intensity up to 25 keV. The total bandwidth of the incident beam can be increased by tapering the gap of the undulator magnets providing nearly 5 keV-wide harmonics. With the fundamental harmonic set well below the energy cutoff of the mirror, Bragg peaks arising across the entire bandwidth of the beamline can be observed in a single Laue pattern.

A new small offset (1 mm) double-bounce Si (111) monochromator was installed and commissioned in the 34-ID-C end station of the Advanced Photon Source. The movable monochromator is located inside a vacuum chamber on top of a granite stage to minimize vibrations. A schematic of the entire assembly is shown in FIG. 1. Air bearings lift the granite table top to permit motors to slide the monochromator crystals in and out of the incident polychromatic x-ray beam. The vacuum chamber is connected to the vacuum of the beamline with stainless steel bellows. By sliding the entire chamber perpendicular to the incident beam direction, the crystals are moved in and out of the pink beam, one can selectively switch from the broadband to the monochromatic beam produced by the Si crystals satisfying the Bragg condition for the energy of our choice. 

\begin{figure}[h]
	\includegraphics[width=0.48\textwidth]{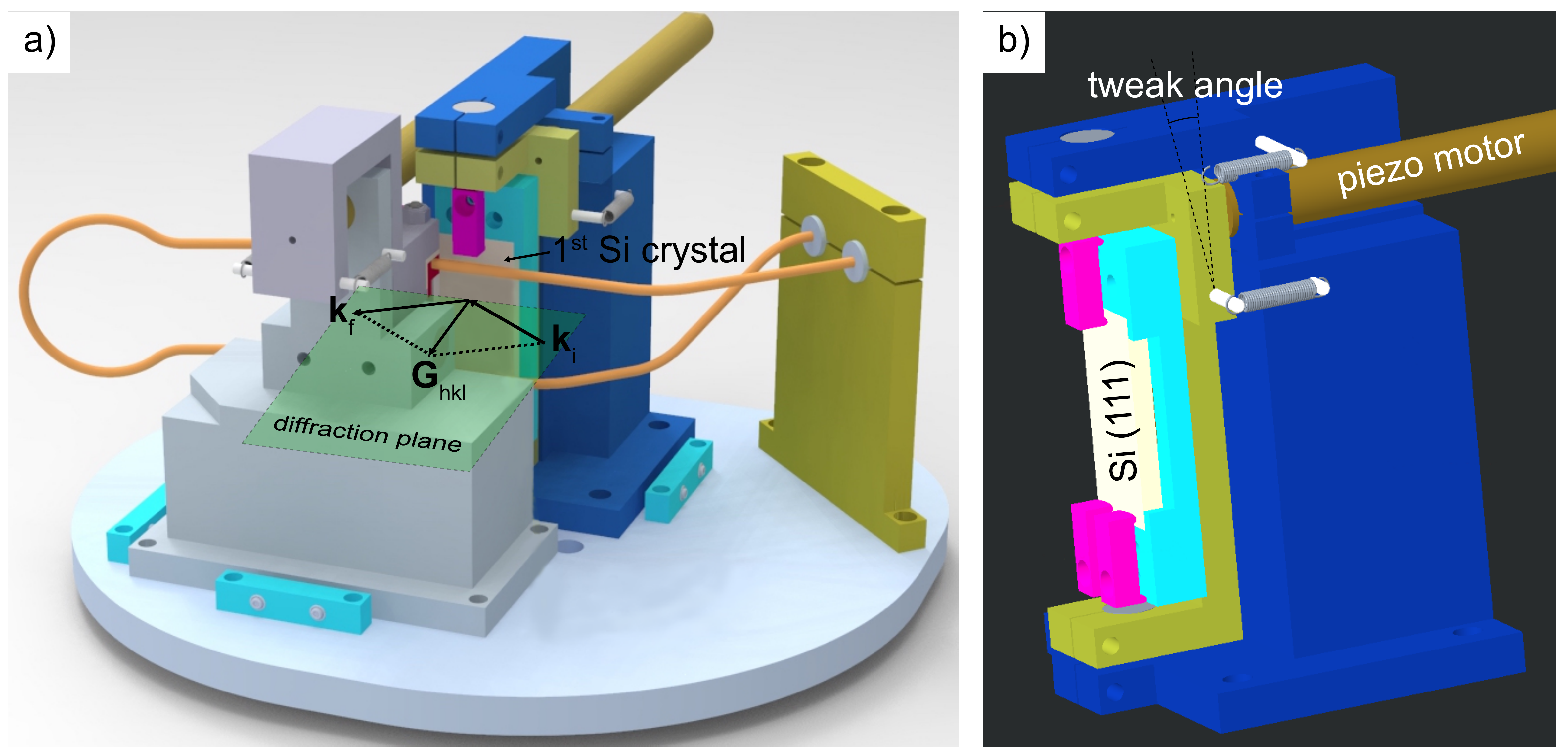}
	\caption{Inside the vacuum chamber. a) Scattering geometry for the incident x-ray beam and the two Si single-crystal wafers. b) The vertical angle (tweak) of the second crystal helps the alignment and optimization of the peak intensity of the exiting monochromatic beam.}
\end{figure}

The monochromator can access x-ray energies from 6 to 20 keV.  Detailed schematics of the monochromator are shown in FIG. 2. The configuration is a standard horizontally reflecting synthetic channel cut. The Si (111) oriented crystals are mounted upon a single rotation table.  Precision alignment of the crystal planes is afforded by two precision orientations, crystal one having a fine rotation about the beam axis ($\chi$) and crystal two having a fine rotation about its Bragg angle (tweak). In FIG. 2a) the scattering geometry of the broadband beam with Si crystals is shown. The diffraction plane is depicted by the transparent green rectangle along with the incident $\mathbf{k}_i$, diffracted beam $\mathbf{k}_f$, and the diffraction vector $\mathbf{G}_{hkl}$. FIG. 2b) shows in greater detail the design of the individual crystal holders. The overall design goal was to have only piezo actuated in-vacuum motions with all long range, stepper-based motors mounted external to the vacuum chamber. 

This both reduces cost and the inevitable thermal drifts within vacuum caused by power cycling of stepper motors. The Bragg angle of the monochromator is actuated by a linear stage mounted parallel to the x-ray direction driving a sine bar that is coupled to the crystal table through a stiff metal strap. The position of the sine bar is monitored with an optical encoder. In addition, the second crystal of the monochromator is heated to increase the lattice constant of the Silicon and modify the exit angle of the diffracted beam to keep the monochromatic beam spatially aligned and coincident to the pink beam in the final focusing optics \cite{RN3011}. This ensures that the apparent source of both the monochromatic and pink beams appear in the same position in the final focusing optics of the experimental instrument. So, one can switch from pink to monochromatic beam with confidence that the focused x-ray beam is on the sample position of the sample. Beamline 34-ID-C uses bendable Kirkpatrick-Baez mirrors as the final focusing optic \cite{RN3035}.  The typical spot size is 500 nm horizontal by 700 nm vertical on the center of rotation of the diffractometer 7 cm downstream of the end of the horizontally focusing mirror.

\subsection{Indexation of Laue patterns}

To demonstrate the capability of indexing Laue patterns from arbitrarily oriented nanocrystals we used single-crystal sub-micron Au particles grown on Si (001) substrates. For the collection of Laue patterns, the experimental geometry depicted in FIG. 3 was used. 
\begin{figure}[h]
	\includegraphics[width=0.35\textwidth]{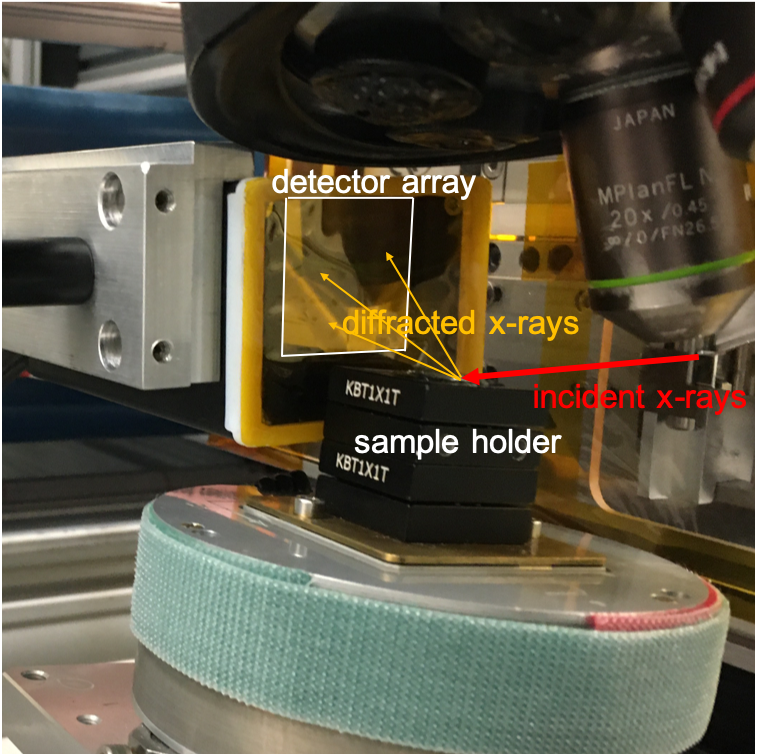}
	\caption{The experimental setup for collecting Laue patterns from Au nanocrystals at 34-ID-C end station of the Advanced Photon Source. The detector is placed 26 mm away from focus at an angle of 90$^\circ$. The red arrow shows the pink beam, which is incident on the sample at 10º angle and the orange arrows depict the trajectories of the diffracted x-rays that contribute to the measured Bragg peaks of the Laue pattern.}
\end{figure}
An Amsterdam Scientific Instruments Timepix QTPX-262k detector with GaAs sensor was mounted 26 mm away from the center of rotation of the diffractometer. This detector is a single-photon counting pixel array detector (512$\times$512 pixels, 55$\times$55 $\mu\mathrm{m}^2$ pixel size). Laue patterns were collected in the plane above and inboard of the incident beam \cite{RN3012}. The active area of the detector is 28.4$\times$28.4 mm$^2$ and at the given distance covers a solid angle of 57.3$^\circ$. While the distance of the detector allows for the collection of a large number of peaks from single-crystal films, smaller crystal objects such as half micron Au nanoparticles do not always produce large enough number of peaks to allow for successfully indexing the pattern.

An example of an indexed Laue pattern is shown in FIG. 4. The Bragg reflections from the Si substrate are five orders brighter than the peaks from one Gold nanoparticle. A strategy was developed to help identify collections of dim peaks. 
\begin{figure}[h]
	\includegraphics[width=0.4\textwidth]{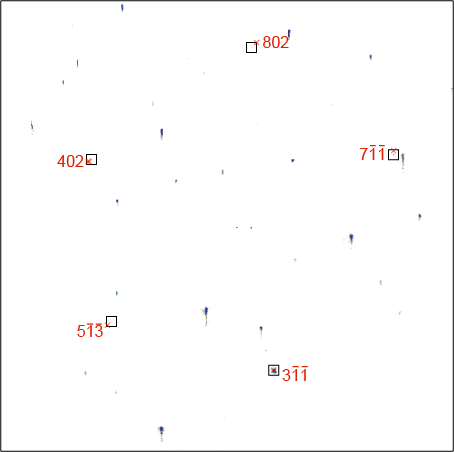}
	\caption{Indexed Laue pattern from a single-crystal Au nanoparticle on a Si substrate. The black squares indicate the location of the Au peaks and the red labels correspond to the Miller indices of the indexed reflections. The bright and vertically elongated peaks originate from the Si substrate and were not considered.}
\end{figure}
We found that repeatedly moving the sample 5 $\mu$m perpendicular to the beam, so the nanoparticle is in and out of the pink beam, we could observe the appearance and disappearance of the corresponding set of reflections. The indexation of the identified Bragg peaks seen on the Laue pattern in FIG. 4 originating from the Au nanocrystal was done with the \textit{LaueGo} package \cite{RN687, RN3014, RN3011}. \textit{LaueGo} uses a plane-search algorithm to identify the conics in the measured Laue pattern in reciprocal space \cite{RN3018, RN3017, RN3013}.

The package takes as input the crystal structure space group of the material, the unit cell dimensions and angles between the different crystallographic axes, the physical dimensions of the active area of the detector, the number of pixels, and the vectors $\mathbf{p}$ and $\mathbf{r}$, which describe the position of the center of the detector with respect to the sample-beam interaction point for a specific experimental geometry. The vector $\mathbf{p}$ consists of the lateral offset of the center of the detector and the distance of the detector from the origin. The vector $\mathbf{r}$ is the Rodriguez rotation vector, which is applied on $\mathbf{p}$ to point at the coordinates of the center of the detector during the measurement in three-dimensional space. During our measurements the center of the detector was found to be displaced a few millimeters from the nominal position in the case of only an applied rotation. We determined the $\mathbf{p}$ vector to be $\mathbf{p} = [13.77, -0.55, -26.46]$ expressed in millimeters. 
\begin{figure}[h]
	\includegraphics[width=0.33\textwidth]{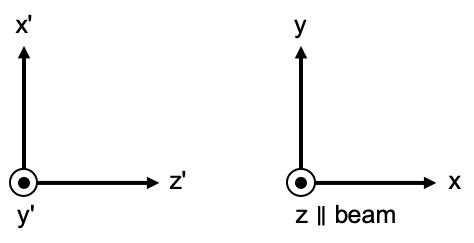}
	\caption{Front view of the experimental geometry depicting the laboratory reference frame on the right and the detection frame on the left. The detector plane is parallel to the $x'y'$ plane.}
\end{figure}
While determining the values of $\mathrm{p}$ relies on accurately measuring the detector to sample distance down to a millimeter, determining the rotation vector is done mathematically. The schematic in FIG. 5 shows the front view of the detection geometry used at 34-ID-C. The incoming x-ray beam is parallel to the z axis and incident to the sample, at the origin of the laboratory reference frame seen on the right of FIG. 5. The surface of the pixel array is laying on the $xy$ plane of the detector frame of reference, seen on the left of FIG. 5. The chips are spatially separated creating a cross-shaped area of blank pixels. Since the detector is rotated by 90$^\circ$, its vertical axis y is parallel to the $z$ axis, with the $z$ axis being the axis of the x-ray wavefront propagation. 

To determine the rotation vector, we need to first calculate the rotation matrix $\mathrm{R}_\mathrm{M}$. The basis vectors $(x, y, z)$ of the laboratory frame are related to the vectors $(x, y, z)$ of the detection frame through the following relations:
\begin{equation}\begin{matrix}
	x=z' \\
	y=x' \\
	z=y'.	
\end{matrix}\end{equation}

Thus, one can derive the rotation matrix by putting the vectors of the detection frame in rows: 
\begin{equation}
\left(\begin{matrix} x\\y\\z\end{matrix}\right)=\mathrm{R}_\mathrm{M}\left(\begin{matrix} x'\\y'\\z'\end{matrix}\right) \Rightarrow \left(\begin{matrix} x\\y\\z\end{matrix}\right) = \left(\begin{matrix} 0&0&1\\1&0&0\\0&1&0\end{matrix}\right)\left(\begin{matrix} x'\\y'\\z'\end{matrix}\right)
\end{equation}
where $\mathrm{R}_\mathrm{M}$\ is the rotation matrix. Using the rotation matrix and the Rodriguez rotation formula \cite{RN3016} we determined the rotation vector to be:
\begin{equation}
	\mathbf{r} =\left(\begin{matrix} \mathrm{1.2}\\ \mathrm{1.2}\\ \mathrm{1.2}\\ \end{matrix}\right).
\end{equation}

Providing the above information that accurately describes the experimental geometry and the crystal structure of Au (crystal space group $Fm\bar{3}m$, lattice constant 4.0782 $\mathrm{\AA}$ we were able to index the Laue pattern shown in FIG. 4 with root mean square error of 0.2$^\circ$. The relatively large error can be further reduced, to the order of 10$^{-4}$ with energy calibration. This part requires the precise determination of the energy of at least one of the experimentally observed reflections on the Laue pattern. However, for our purposes, 0.2$^\circ$ accuracy is sufficient since the coherent diffraction extends for nearly one degree about a given Bragg peak. The final result of the indexation contains a list of the indexed peaks with their Miller indices, energies, pixel coordinates on the detector, and the error with respect to the unstrained unit cell. The results are summarized in TABLE \ref{table}. 
\begin{table}[h]
	\begin{center}
	\caption{\label{table} Miller indices of the indexed reflections from the Au nanoparticle, their energies, and error. \\}
		\begin{tabular}{l|c|r} 
			\textbf{(hkl)} & \textbf{Energy} [keV] & \textbf{Error} [$\circ$] \\
			$(3\bar{1}1)$ & 8.21 & 0.07 \\
			$(7\bar{1}1)$ & 14.28 &	0.16 \\
			$(402)$	 & 8.86 & 0.14 \\
			$(802)$	& 15.14	& 0.48 \\
			$(5\bar{1}3)$ & 14.10 & 0.39 \\
			$(11 1 7)$ & 25.99 & 0.13
		\end{tabular}
\end{center}
\end{table}

The most important information obtained by the indexation of a Laue pattern is the orientation matrix $\mathrm{U}_\mathrm{c}$. The orientation matrix obtained from the Laue pattern of FIG. 4 contains the reciprocal space lattice vectors $\mathbf{a}^\mathrm{*}, \mathbf{b}^\mathrm{*}, \mathbf{c}^\mathrm{*}$ in the crystal frame given in units of inverse nanometers (nm$^{-1}$):
\begin{equation}
	\mathrm{U}_\mathrm{c}=\left(\ \begin{matrix}\mathrm{|}&\mathrm{|}&\mathrm{|}\\\mathbf{a}^\mathrm{*}&\mathbf{b}^\mathrm{*}&\mathbf{c}^\mathrm{*}\\\mathrm{|}&\mathrm{|}&\mathrm{|}\\\end{matrix}\ \right)=\left(\ \begin{matrix}\mathrm{-4.44}&\mathrm{7.68}&\mathrm{-12.59}\\\mathrm{6.88}&\mathrm{-10.55}&\mathrm{-8.86}\\\mathrm{-13.04}&\mathrm{-8.18}&\mathrm{-0.38}\\\end{matrix}\ \right).
\end{equation}
While the orientation matrix $\mathrm{U}_\mathrm{c}$, contains the necessary information of how the crystal is oriented in reciprocal space, it does not consist of a convenient representation in order to easily navigate from one Bragg reflection to another. Thus, we will translate this information to the real space coordinate system of the laboratory, and more specifically, to a set of two real-space vectors, one with the out-of-plane direction of the scattering planes and one with one in-plane direction, denoted by $\mathbf{H}_\mathrm{\bot}$ and $\mathbf{H}_\mathrm{\parallel}$, respectively. To translate the orientation matrix $\mathrm{U}_\mathrm{c}$ into the two vectors, we first apply the relative rotation matrices $\mathrm{R}_\mathrm{\theta}$, $\mathrm{R}_\mathrm{\phi}$, and $\mathrm{R}_\mathrm{\chi}$ based on the values of the angles $\mathrm{\theta}$, $\phi$, and $\mathrm{\chi}$ of the goniometer. This gives us a new matrix expressed in the laboratory frame: 
\begin{equation}\mathrm{U}_\mathrm{L}=\mathrm{R}_\mathrm{\theta}\mathrm{R}_\mathrm{\phi}\mathrm{R}_\mathrm{\chi}\mathrm{U}_\mathrm{c}\end{equation}
with
\begin{equation}\mathrm{R}_\mathrm{\theta} = \left(\begin{matrix} \cos\theta & 0 & -\sin\theta \\ 0 & 1 & 0 \\ \sin\theta & 0 & \cos\theta\end{matrix}\right),\end{equation}
\begin{equation}\mathrm{R}_\mathrm{\phi} = \left(\begin{matrix} 1 & 0 & 0 \\ 0 & \cos\phi & -\sin\phi \\ 0 & \sin\phi & \cos\phi \end{matrix}\right),\end{equation}
and
\begin{equation}\mathrm{R}_\mathrm{\chi} = \left(\begin{matrix} \cos\chi & -\sin\chi & 0 \\ \sin\chi & \cos\chi & 0 \\ 0 & 0 & 1\end{matrix}\right).\end{equation}

The $\mathbf{H}_\mathrm{\bot}$, $\mathbf{H}_\mathrm{\parallel}$ vectors are obtained if we multiply the unitary vectors $\hat{\mathrm{x}}$, and $\hat{\mathrm{y}}$ in the laboratory frame with the transpose of the $\mathrm{U}_\mathrm{L}$ matrix ($\mathrm{U}_\mathrm{L}^\mathrm{T}$):
\begin{equation}{\mathbf{H}_\mathrm{\bot}\mathrm{=U} }_\mathrm{L}^\mathrm{T}\hat{\mathrm{x}}\end{equation}

\begin{equation}{\mathbf{H}_\mathrm{\parallel}\mathrm{=U} }_\mathrm{L}^\mathrm{T}\hat{\mathrm{y}}.\end{equation}

The calculation gives us for the experimentally determined orientation matrix, $\mathrm{U}_\mathrm{c}$, the following vectors:
\begin{equation}\mathbf{H}_\mathrm{\bot}=\left(\begin{matrix}\mathrm{8.95}&\mathrm{-8.86}&\mathrm{-8.88}\\\end{matrix}\right)^\mathrm{T}\end{equation}
which means that the scattering planes are $\left(\mathrm{1}\bar{\mathrm{1}}\bar{\mathrm{1}}\right)$ as illustrated in FIG. 4, and
\begin{equation}\mathbf{H}_\mathrm{\parallel}=\left(\begin{matrix}\mathrm{-5.59}&\mathrm{6.94}&\mathrm{-12.57}\\\end{matrix}\right)^T\end{equation}
equivalent to a $\left[\bar{\mathrm{1}}\mathrm{1} \bar{\mathrm{2}}\right]$ crystallographic direction.
\begin{figure}[h]
	\includegraphics[width=0.33\textwidth]{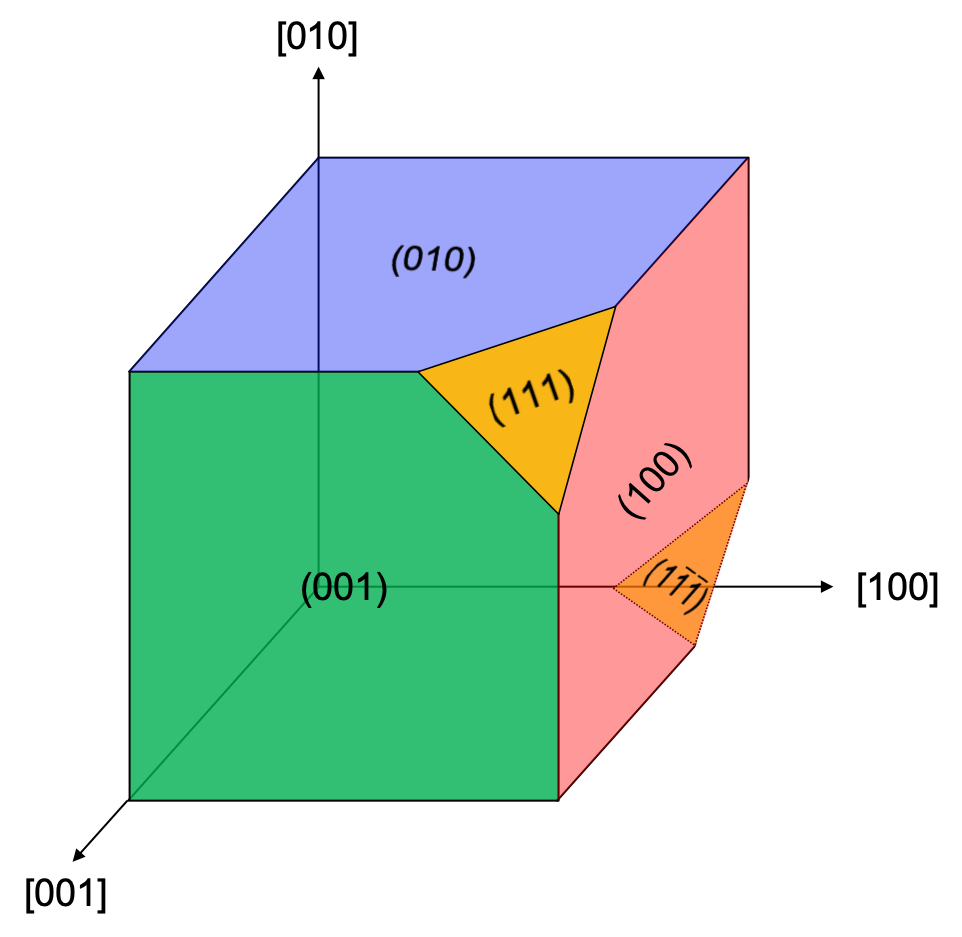}
	\caption{A schematic depicting the crystallographic directions in square brackets and families of atomic planes in parentheses of a simple cubic crystal system.}
\end{figure}

The $\mathbf{H}_\mathrm{\bot}$, $\mathbf{H}_\mathrm{\parallel}$ vectors are then inserted in \textit{Spec}, allowing to move the detector arm and the rotation stages by simply giving as input the Miller indices of the desired Bragg reflection \cite{RN3015}. The schematic below (Fig. 6) illustrates how one can navigate reciprocal space in order to reach different Bragg peaks given the two orientation vectors $\mathbf{H}_\mathrm{\bot}$, $\mathbf{H}_\mathrm{\parallel}$. In particular, possessing the out-of-plane vector of the diffracting planes and one in-plane vector offers all the necessary information for reaching the other Bragg peaks. For example, FIG. 6 shows that by knowing that the family of $\left(\mathrm{1}\bar{\mathrm{1}}\bar{\mathrm{1}}\right)$ planes is contributing to scattering, we can rotate our crystal as needed in $\mathrm{\theta}$, $\phi$, and $\mathrm{\chi}$ in order to measure a $(001)$ reflection \cite{RN3015}.

\section{Calculation of the strain tensor}

Using the indexation result from Laue pattern of FIG. 4 we found four Bragg reflections originating from the same single-crystal Au nanoparticle, simply by typing the desired Miller indices. FIG. 7 shows slices through the measured 3D x-ray diffraction patterns at the maximum intensity for the $\left(\mathrm{11}\bar{\mathrm{1}}\right)$, $\left(\bar{\mathrm{1}}\bar{\mathrm{1}}\bar{\mathrm{1}}\right)$, $\left(\mathrm{1}\bar{\mathrm{1}}\mathrm{1} \right)$, and $\left(\mathrm{2}00\right)$ Bragg reflections obtained with the monochromatic beam at 9 keV. 
\begin{figure}[h]
	\includegraphics[width=0.4\textwidth]{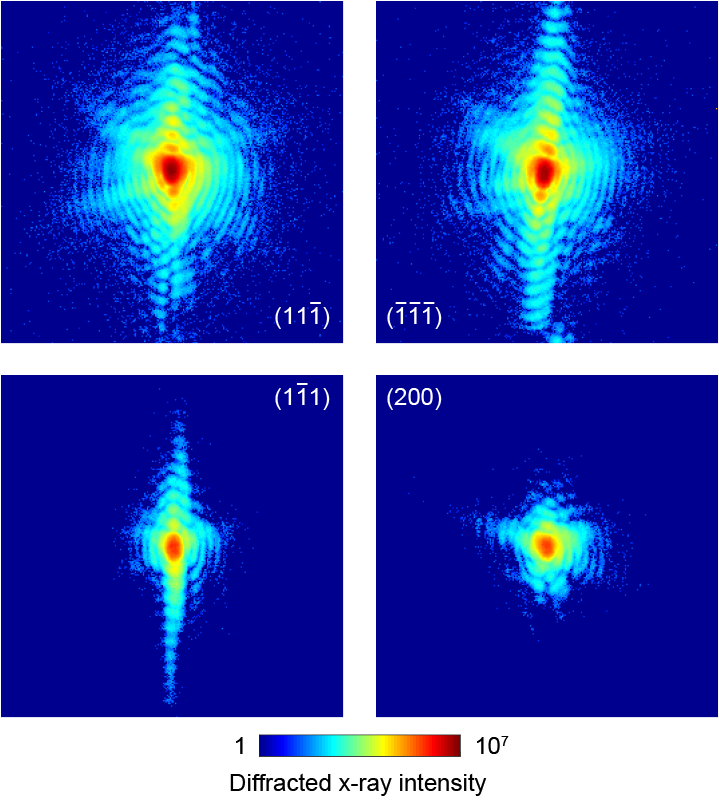}
	\caption{Measured x-ray diffraction patterns from a single Au nanoparticle for four different Bragg reflections. The patterns were taken at the intensity maximum of the rocking curve during the BCDI measurement.}
\end{figure}
The patterns demonstrate high visibility fringes that extend far in reciprocal space from the sharp facets of the Au particle, features that assist obtaining high resolution BCDI reconstructions \cite{RN3024}. For all the collected datasets from the four Bragg reflections, we reconstructed the 3D crystal electron density shown in FIG. 8. 
\begin{figure}[h]
	\includegraphics[width=0.36\textwidth]{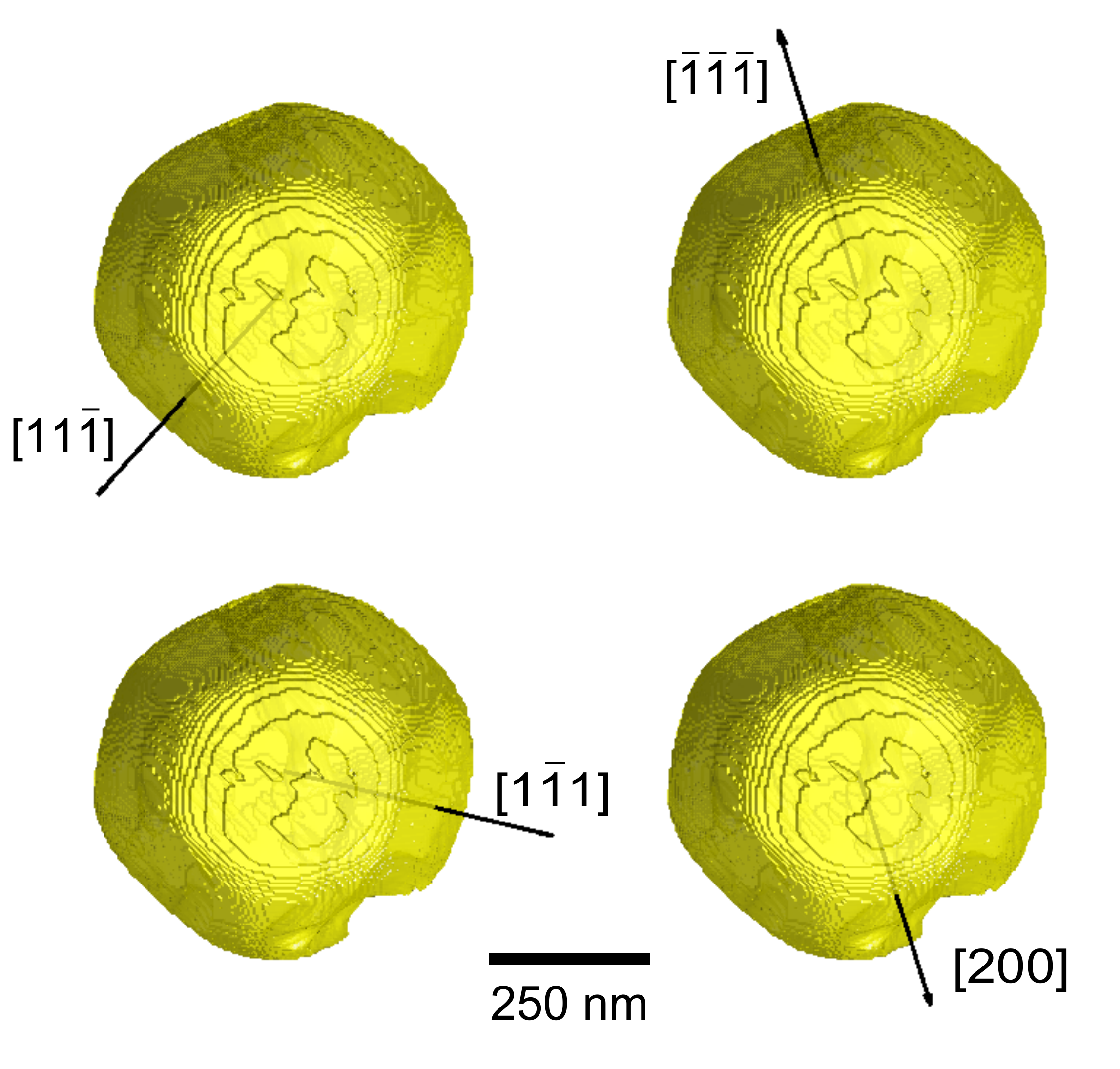}
	\caption{Three-dimensional reconstructions of the measured Au nanocrystal from four different Bragg reflections.}
\end{figure}
The shape, dimensions, and surface features observed at 0.4 threshold isosurface of the Au nanocrystal are consistent among the reconstructions from the different reflections.

In addition to the 3D crystal electron density, we plot the arrows indicating the crystallographic orientation axes for each measurement. The arrow is normal to the $(hkl)$ family of atomic planes, which contributes to the scattering signal. A series of alternating error-reduction and difference map algorithms was used for 620 iterations \cite{RN108} and the phase-retrieval transfer function was employed for the calculation of the resolution of the obtained reconstruction \cite{RN2968, RN410}. We estimated an isotropic resolution for all three dimensions of approximately 15 nm, slightly larger than the voxel size of 11 nm. Comparison of the reconstructed shapes was made with the computational methods previously developed by \cite{RN3015}. 

Having all components of the crystal atomic displacement we can calculate the entire Eulerian strain tensor $\varepsilon_{\mathrm{ij}}$, given by the following relation \cite{RN1676}
\begin{equation}\varepsilon_{ij}=\frac{1}{2}\left(\frac{{\partial u}_j}{{\partial x}_i}+\frac{{\partial u}_i}{{\partial x}_j}\right).\end{equation}

The result of the calculation is plotted in FIG. 8, where slices taken at the center of each of the 3D reconstructions. Each slice depicts the different crystal strain components $\varepsilon_{ij}$ of the strain tensor \cite{RN3019}. The reconstructed slices show in general the presence of small strain fields inside the crystal on the order of 10$^{-4}$, except in the case of the $\varepsilon_{\mathrm{yy}}$ component, for which the bottom part of the Au particle reaches values up to 8$\times10^{-4}$. This is likely due to stress transmitting through the interface of the Au particle with the Si substrate \cite{RN2834, RN2797}.

In comparison with other optical methods for measuring full-field displacements and strains, such as digital image correlation, BCDI provides strain information in 3D allowing the investigation of dislocations or other nanoscale phenomena inside a single grain \cite{RN3029, RN3030}. Additionally, integrating BCDI with atomistic simulations \cite{RN3025} can provide important insight in understanding the role of grain boundaries during damage and failure \cite{RN3032, RN3031}.

\begin{figure}[h]
	\includegraphics[width=0.48\textwidth]{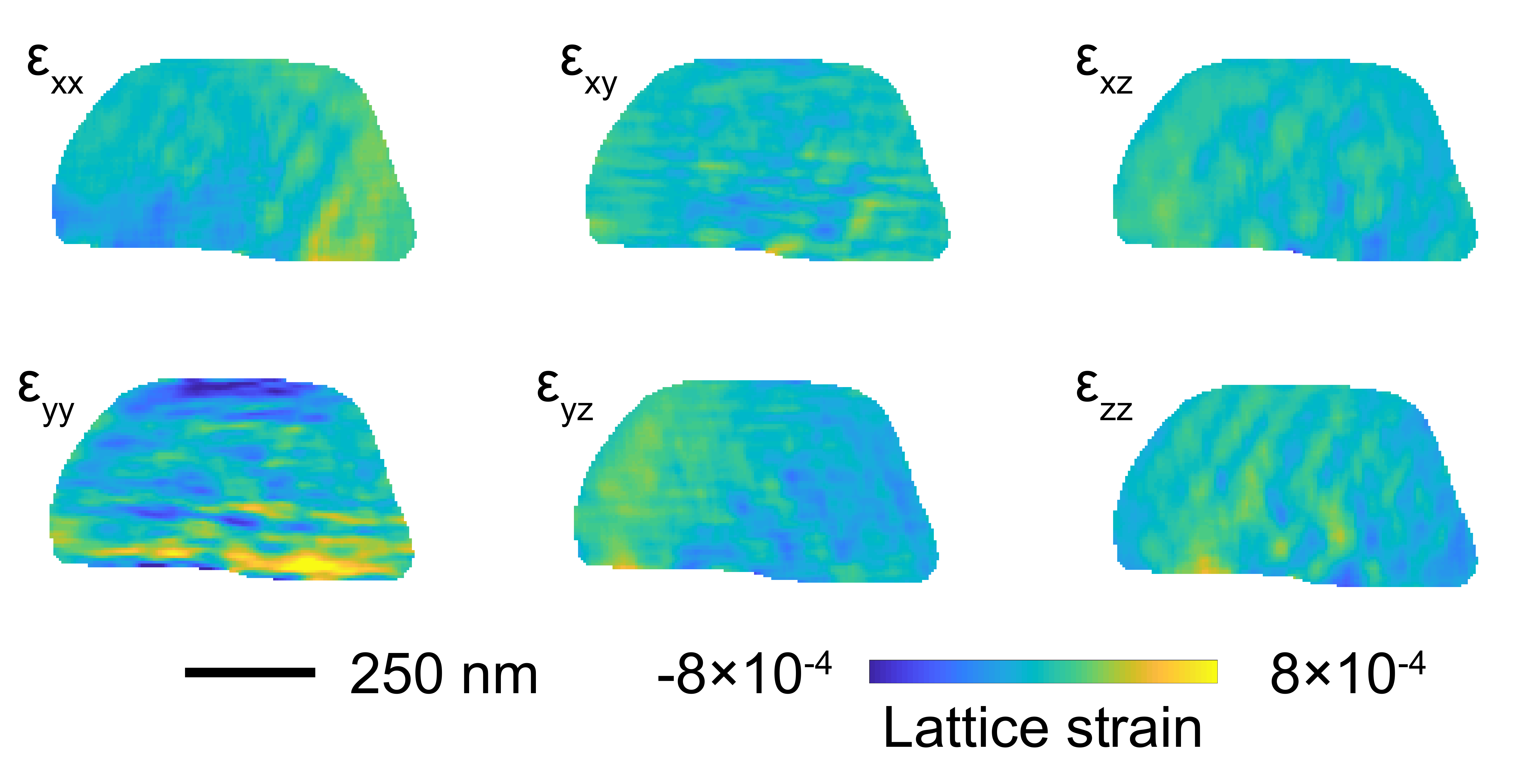}
	\caption{Two-dimensional slices of the strain components taken at the center of the reconstructed volumes of the Au nanoparticle.}
\end{figure}

\section{Conclusions}

The commissioning of a movable x-ray monochromator at 34-ID-C end station of the Advanced Photon Source will change dramatically the user workflow, from sample preparation to measurement strategies, data analysis, and time management during the beamtime. Initially it will require modifying the design of samples since particularly dense specimens make the indexation of Laue patterns a challenging problem. However, strategies for identifying Laue patterns from multiple grains within the beam are being developed. Future developments with potentially the integration of machine learning algorithms to assist the indexation of complex Laue patterns \cite{RN3034}, or 3D reconstructions based on variable-wavelength coherent x rays rather than rocking crystals \cite{RN286, RN3033} will benefit the effort of standardizing each experiment, minimizing the man hours spent finding good signals to run experiments and focusing on the quick analysis and interpretation of numerous datasets. We are also developing a series of procedures, which aim at facilitating finding different reflections using the result of the indexation. While in this work we described the steps currently needed to calculate the in- and out-of-plane orientation vectors, the operations can be done directly after importing the result of the LaueGo package and inserted in \textit{Spec} automatically with the generation of an improved protocol. We believe these developments will accelerate research in different fields such as materials science, condensed matter physics, and chemistry.

\bibliography{main.bib}

\end{document}